# Solar Neutrino Interactions with $^{18}$O in SuperKamiokande

W.C. Haxton and R.G.H. Robertson

*Institute for Nuclear Theory, Box 351550*

*and Department of Physics, Box 351560*

*University of Washington, Seattle, Washington 98195-1550*

(January 23, 2018)

## Abstract

The ratio of $^{18}$O nuclei to electrons in SuperKamiokande is quite small, about 1:5000. However this nucleus provides two unpaired neutrons and a low threshold for solar neutrino absorption, resulting in an unusually large cross section. The $^{18}$O nuclei, together with $^{17}$O and deuterium, will produce in excess of 460 scattered electrons per year with energies above the initial SuperKamiokande threshold of 6.5 MeV, or about 6.7% of the total solar neutrino signal, for an electron neutrino flux scaled to the current SuperKamiokande counting rate. We explore whether this might complicate efforts to extract the neutrino spectrum from elastic scattering events and whether there might be any prospect of combining $^{18}$O and elastic scattering events to separate charged and neutral current interactions. Under current operating conditions the answers are no, as the Fermi, Gamow Teller, and forbidden contributions to the $^{18}$O cross section conspire to produce a nearly isotropic cross section. Thus most of the events are indistinguishable from background. This result is relevant to SuperKamiokande detection of solar antineutrinos, which arise in certain oscillation scenarios.

The SuperKamiokande (SuperK) water Cerenkov detector [1], with an inner fiducial volume of 22.5 kilotons, is in its second year of operations. In this detector the scattering of $^8$B solar neutrinos off electrons produces a sharply forward peaked electron spectrum, permitting these events to be separated from a background that is uncorrelated with the position of the sun. The results of 504 days of running with an electron detection threshold of 6.5 MeV is a counting rate consistent with an undistorted $^8$B electron neutrino flux of $(2.44^{+0.05+0.09}_{-0.05-0.07}) \times 10^6/\text{cm}^2\text{s}$ [5], or 36.8 % of the standard solar model (SSM) prediction of Bahcall and Pinnsoneault (BP95) [3].

From these early results, plus those of the $^{37}$Cl [6], SAGE/GALLEX [7], and Kamioka II/III [2] experiments, a pattern of solar neutrino fluxes emerges that is very difficult to accommodate by modifying the SSM. A more natural explanation is provided by matter enhanced neutrino oscillations [4]. Many regard the solar neutrino problem as the best evidence for massive neutrinos and thus of physics beyond the standard electroweak model.

The $\nu + e$ cross sections for heavy flavor neutrinos is about 1/6 that for electron neutrinos. Thus, if the SuperK discrepancy with the standard solar model is due to flavor oscillations, one expects that all but about 23% of the $^8$B flux has been transformed into muon or tauon neutrinos. One goal of SuperKamiokande is to distinguish this possibility from others, such as a flux comprised entirely of electron neutrinos but reduced by $\sim 0.368$ relative to the standard solar model. This cannot be done from the total $\nu + e$ counting rate, as no explicit flavor separation can be made. However the oscillation scenario that best fits [8] current measurements, the so-called small-angle solution, produces a subtle distortion in the energy distribution of scattered electrons that could be exploited as a signal for new physics. The detection of this distortion requires a precise understanding of the detector's efficiency as a function of scattered electron energy $E_e$. ($E_e$ is the total energy, mass plus kinetic.) The SuperK collaboration has calibrated [5] the detector with electrons from a linac, with the goal of achieving an understanding of the detector's resolution at the level of 1%. Also crucial is the extent to which background rates in the detector can be reduced, as this permits the threshold energy for detecting electrons to be lowered: the distortion is most apparent in the lowest energy bins. In the most recent runs, the data-taking threshold has been lowered to about 5 MeV.

As one of us discussed [9] some time ago in connection with the Kamiokande II experiment, there is a second solar neutrino reaction that will produce energetic neutrinos in a water Cerenkov detector. Electron neutrinos will interact by the charged current reaction $(\nu_e, e)$ with the isotope $^{18}$O (abundance 0.204%), as well as with $^{17}$O (0.038%) and deuterium (0.015%). As will be discussed in more detail below, the $^{18}$O cross section averaged over the normalized $^8$B spectrum is exceptionally large,

$$\langle \phi(^8\text{B})\sigma \rangle = 5.35 \times 10^{-42} \text{cm}^2, \tag{1}$$

and exceptionally hard, $\langle E_e \rangle = \langle E_\nu \rangle$ - 1.37 MeV. Thus $\sim 70\%$ of the electrons produced are above the 6.5 MeV threshold initially employed in SuperK. The net result, assuming an electron neutrino flux of $\phi(^8\text{B}) = 0.368\ \phi_{SSM}(^8\text{B})$, is about 460 detectable events per year (1.27 events/day) arising from nuclei, 92.6% of which are due to $^{18}$O. This can be compared to the $\nu + e$ counting rate of about 19 events/day. Thus, despite the 1:5000 ratio between $^{18}$O nuclei and electrons in water, the nuclear contribution to the solar neutrino capture rate is not insignificant, comprising 6.7% of the total signal.

This immediately raises the two questions we will attempt to answer in this paper. First, the forward-peaked $\nu + e$ spectrum, which is spread over a significant solid angle due to the finite resolution of the detector ($\sim \pm 28°$ at $E_e = 10$ MeV), must be extracted from a larger but very nearly isotropic background of $\sim 200$ events/day. The $^{18}$O events arise from a combination of backward-peaked Gamow-Teller (GT) ($\sim 1 - \frac{1}{3}\cos\theta_{\nu e}$) and forward-peaked Fermi ($\sim 1 + \cos\theta_{\nu e}$) nuclear transitions and presumably account for a small fraction ($\sim 0.7\%$) of the events presently identified as background. The resulting combination of GT and Fermi transitions is almost isotropic, but with the addition of recoil-order terms becomes increasing backward peaked for the largest $E_e$. Could the residual anisotropy, if not taken into account, affect the analysis of $\nu - e$ events? This question is important because there is slight residual structure in SuperK's observed electron spectrum as a function of $E_e$, relative to that predicted for an undistorted $^8$B neutrino flux [5]. Second, is there any strategy - perhaps involving some combination of future lower background rates in SuperK, careful angular analysis, and study of annual variations - that might allow identification of the $^{18}$O events? This would be an exciting development as a measurement of the $\nu_e$ flux that could distinguish a reduction factor of 0.23 from 0.368 would provide an important, direct test of oscillations.

The ground state of $^{18}$O has $J^\pi T = 0^+1$. Neutrinos above the threshold energy $E_\nu = 1.655$ MeV can excite the charged-current reaction $^{18}$O$(\nu,e)^{18}$F. The cross section for $^8$B neutrinos is dominated by three transitions of known strength. The GT transition to the $1^+0$ ground state of $^{18}$F is fixed by the inverse $\beta$ decay rate. The Fermi transition to the 1.04 MeV $0^+1$ first excited state and the GT transition to the 1.70 MeV $1^+0$ second excited state can be determined from the analog $^{18}$Ne $\to^{18}$F $\beta$ decay partial rates. The former is consistent with an allowed matrix element of $|M_F|^2 = 2$. Table I, which summarizes the matrix elements we use, shows that the ground and first excited state transition are very strong, with log(ft) = 3.08 and 3.47, respectively. The ground, 1.04 MeV, and 1.70 MeV states carry 79.5%, 19.1%, and 1.4% of the $^8$B neutrino absorption cross section.

Other allowed transitions exist, but at substantially higher excitation energies. A simple $sd$ shell model calculation with the Brown-Wildenthal [10] interaction predicts very weak GT transitions to excited states at 7.26 and 10.23 MeV, with stronger transitions populating states at and above 13 MeV. The net result is an increase in the cross section of 0.0002%. However it is known, in contradiction to this naive picture, that $4p2h$ excitations occur at relatively low excitation energies in $^{18}$F: additional $1^+$ states begin at 3.724 MeV. Thus to generate a more conservative estimate of the contributions of such states, we placed the entire strength carried by higher shell model states, $|M_{GT}|^2 = 0.67$, at this threshold energy. (In making this estimate we follow the usual $sd$ shell model procedure of taking $g_A^{eff} = 1$.) The resulting contribution to the cross section is $\sim 1\%$. We conclude that unknown GT transitions to excited states above 3.724 MeV cannot make an appreciable contribution to the $^8$B neutrino cross section. As a solar neutrino target, $^{18}$O is unusual in that the cross section is determined almost entirely by measured $\beta$ decay rates, and thus is rather precisely known. [Forward-angle (p,n) measurements for $^{18}$O and $^{17}$O were recently reported [11]. The extraction of GT strengths from these data are in progress. However, the cross section for the state at 3.7 MeV in $^{18}$O appears to be similar to that for the state at 1.7 MeV, which has $|M_{GT}|^2 = 0.21$. Thus the 1% estimate for higher lying GT strength is likely a reliable bound.]

As the cross section depends on just three transitions, the double differential cross section averaged over $^8$B neutrinos becomes, in the allowed approximation,

$$\langle \frac{d\sigma_{18}}{d\cos\theta_{\nu e} dE_e} \rangle = \frac{G_F^2 \cos^2\theta_C}{2\pi} k_e E_e F(Z=9, E_e) \times$$
$$\left( (1 - \frac{1}{3}\beta_e \cos\theta_{\nu e})(5.12\phi_\nu(E_\nu^{gs}) + 0.21\phi_\nu(E_\nu^{1.70})) + 2.00(1 + \beta_e \cos\theta_{\nu e})\phi_\nu(E_\nu^{1.04}) \right). \quad (2)$$

where $E_\nu^{gs} = E_e + 1.144$ MeV, $E_\nu^{1.04} = E_e + 2.186$ MeV, and $E_\nu^{1.70} = E_e + 2.844$ MeV. Here $k_e$ is the magnitude of the electron's momentum, $\beta_e = k_e/E_e$, $\theta_{\nu e}$ is the angle between the scattered electron and incident neutrino, $G_F$ and $\theta_C$ are the Fermi coupling and Cabibbo angle, and $F(Z, E_e)$ is the standard Fermi-function correction [12] for the Coulomb distortion of the electron wave function in the field of the daughter nucleus $^{18}$F. Thus one sees, for a given $E_e$, that each contributing transition samples a different portion of the normalized electron neutrino flux density $\phi_\nu(E_\nu)$. As a result, even in lowest order, the angular behavior has some dependence on $E_e$.

As the coefficients of the GT and Fermi terms in Eq. (2) are approximately in the ratio 3:1, it is apparent the relatively strong angular dependences of the separate Fermi and GT contributions tend to cancel when summed. The net result is that corrections to the allowed approximation, while increasing the total cross section by only 3.5%, must be included in calculations of the angular variation. The leading-order contribution is due to weak magnetism, as this generates an interference term between an allowed GT matrix element and the isovector magnetic moment, thus producing a contribution to the cross section proportional to the three-momentum transfer q. Our results, which include the full momentum dependence of the operators, can be well approximated by adding to the factor

$$(1 - \frac{1}{3}\beta_e \cos\theta_{\nu e})(5.12\phi_\nu(E_\nu^{gs}) + 0.21\phi_\nu(E_\nu^{1.70})) \quad (3)$$

in Eq. (2) the additional term

$$(1 - \cos\theta_{\nu e})(5.12\alpha_1(E_e)\phi_\nu(E_\nu^{gs}) + 0.21\alpha_2(E_e)\phi_\nu(E_\nu^{1.70})). \quad (4)$$

Over the kinematic range of interest, 5 MeV $\lesssim E_e \lesssim$ 12 MeV, $\alpha_1(E_e) \sim 0.0053 E_e$ and $\alpha_2(E_e) \sim 0.010 E_e$, to an accuracy of $\sim$ 20%. The approximate dependence of $\alpha_i$ on $E_e$ is consistent with the weak magnetism-GT interference dominating forbidden corrections. More exact values for $\alpha_1(\alpha_2)$ are 0.029 (0.057), 0.038 (0.073), 0.045 (0.085), 0.050 (0.094), and 0.052 (0.099) at $E_e$ = 4, 6, 8, 10, and 12 MeV, respectively.

We evaluate the much smaller contributions of $^{17}$O and deuterium in the same way. For $^{17}$O, neutrinos above 2.761 MeV can excite the mixed Fermi-GT transition to the analog ground state of $^{17}$F. The corresponding allowed matrix element $|M_F|^2 + |M_{GT}|^2 = 2.69$ can be deduced from the $^{17}$F half life. This leads to a cross section averaged over the normalized $^8$B spectrum of

$$\langle \phi(^8\text{B})\sigma \rangle = 1.53 \times 10^{-42} \text{cm}^2. \quad (5)$$

where the effects of weak magnetism and other forbidden corrections are responsible for 4% of the total. There is a large gap to the next state, the $3/2^+$ level at 5.00 MeV in $^{17}$F, that can be populated in an allowed transition. If one identifies this as the single-particle $1d_{5/2} \to 1d_{3/2}$ transition of the naive shell model, then it leads to an increase in the $^{17}$O cross section of 5.5%. However virtually none of this strength is relevant to SuperK, as it produces events below threshold: only $^8$B neutrinos with $E_\nu$ above 13.5 MeV can contribute while producing scattered electrons above 6.5 MeV. Thus only the ground state transition is important for SuperK. It follows that $E_e = E_\nu - 2.25$ MeV for reactions off $^{17}$O; 58.9% of $^8$B neutrino reactions to the ground state of $^{17}$F produce an electron with $E_e$ above 6.5 MeV. Because of the lower cross section and lower isotopic abundance, the $^{17}$O contribution to the total SuperK counting rate off oxygen is only 0.05/day, or 4.3% of the nuclear signal. The double differential cross section averaged over $^8$B neutrinos is

$$\langle \frac{d\sigma_{17}}{d\cos\theta_{\nu e} dE_e} \rangle = \frac{G_F^2 \cos^2\theta_C}{2\pi} k_e E_e F(Z=9, E_e) \phi_\nu(E_\nu^{gs})(2.69 + 0.44\beta_e \cos\theta_{\nu e}) \quad (6)$$

in the allowed approximation, where $E_\nu^{gs} = E_e + 2.250$ MeV. Thus this contribution is modestly forward peaked.

The charged current cross section for the breakup of deuterium, which proceeds by GT transitions, is slightly smaller than that of O$^{17}$

$$\langle \phi(^8\text{B})\sigma \rangle = 1.15 \times 10^{-42} \text{cm}^2. \quad (7)$$

As the abundance of deuterium is 0.015%, the ratio of deuterium to O$^{17}$ nuclei in water is 0.79. Thus deuterium accounts for 3.1% of the nuclear signal, or 0.04 counts/day. The differential cross section, including forbidden corrections, has been discussed in detail elsewhere [13]: the angular distribution is $\sim 1 - \frac{1}{3}\beta_e \cos\theta_{\nu e}$.

We can now proceed to answer the questions posed in the introduction. In Table II we present SuperK rates, in units of counts/day/MeV, for an undistorted $^8$B neutrino flux reduced to 0.368 of the BP95 SSM value, in order to mimic current counting rates. The size of the contribution from $^{18}$O (and $^{17}$O and deuterium) is significant, averaging 6.7% of the $\nu - e$ elastic scattering rate for $E_e$ above 6.5 MeV and exceeding 10% for $E_e \sim 10$ MeV. The nuclear rate peaks at $\sim 0.3$ counts/day/MeV for $E_e \sim 8$ MeV. Also shown, for comparison, are approximate backgrounds rates found in SuperK under its present operating conditions. (Note that the background rates, as measured quantities, are known as a function of the apparent electron energy $E_e^A$, the result of folding the true energies of events with the detector resolution. All of the calculated rates in this paper are presented for $E_e$, not $E_e^A$.)

The angular dependence of the nuclear contribution, determined from Eqs. (2) and (4), is shown is Fig. 1. It is immediately apparent that the angular asymmetry is quite modest in energy bins where the nuclear counting rate is large, a result of a fortuitous (from one perspective) cancellation between the forward-peaked Fermi and backward-peaked GT contributions. The numerical results for the angular dependence of the counting rate are well approximated by $1 - B(E_e)\cos\theta_{\nu e}$. The derived values of $B(E_e)$ are given in Table II. Note that the cross section evolves from almost exact isotropy at $E_e = 5$ MeV to one with a GT profile at the largest recoil energies.

While the precise impact of the residual asymmetry will depend on details of the experimental analysis, one can make a reasonable estimate of its effect. Because of the angular resolution of SuperK of $\sim 28°$, the strongly peaked $\nu - e$ cross section spreads, producing a tail that is apparent over most of the forward hemisphere. In effect, then, the event rate in the backward hemisphere determines the background subtraction. Therefore we can estimate the consequences of a naive background analysis - one that ignores the asymmetry under discussion - by calculating the difference of the $^8$B neutrino cross sections averaged over backward and forward hemispheres,

$$\langle \frac{d\sigma}{dE_e} \rangle^{b-f} = \int_{-1}^{0} \langle \frac{d\sigma}{d\cos\theta_{\nu e} dE_e} \rangle d\cos\theta_{\nu e} - \int_{0}^{-1} \langle \frac{f\sigma}{d\cos\theta_{\nu e} dE_e} \rangle d\cos\theta_{\nu e} \tag{8}$$

and then comparing the corresponding oxygen rate asymmetry, $R^{b-f}$, to the elastic scattering rate, $R(\nu-e)$. The results are given in Table II. One concludes that a failure to account for the angular asymmetry associated with solar neutrino interactions with nuclei would have led to an underestimation of the extracted elastic scattering rate of $\sim$ 0.1-1.6%, depending on the value of $E_e$, with the average effect for all events above $E_e = 6.5$ MeV being 0.36%. This can be compared to the overall systematic error estimate after 374 days of +3.7%/-2.5%. Thus, the effect is quite small in the total cross section, and remains modest for all energy bins. The current SuperKamiokande backgrounds appear to be independent of angle within errors.

The existence of an appreciable charged-current rate in SuperK is tantalizing because a comparison with the elastic scattering rate could provide a means of distinguishing electron neutrinos from heavy flavor neutrinos. The two possibilities for separating these events from a much larger background are the angular dependence - which we have seen is weak - and the 7% variation of signal with the earth's orbital motion. However, a quick inspection of Table II shows that these signals are inadequate given current background rates.

The presence of an isotropic background dilutes the effect of the angular distribution of the nuclear contributions. The distribution becomes $1 - B'(E_e)\cos(\theta_{\nu e})$, where

$$B'(E_e) = B(E_e) \frac{R(^{18}\text{O} +^{17}\text{O} +^2\text{H})}{R_T(E_e)}, \tag{9}$$

where $R_T(E_e)$ is the total number of events (background plus signal) at energy $E_e$. [The relatively small elastic scattering signal $R(\nu - e, E_e)$ has also been included with the background, although its strong angular dependence decouples it somewhat from terms that are isotropic or linear in $\cos\theta_{\nu e}$.]

If a large isotropic background underlies a signal linear in $\cos\theta_{\nu e}$, it may readily be shown that the fractional uncertainty in $B'(E_e)$ is given by

$$\sigma(B') = \sqrt{\frac{3n^2}{(n+1)(n-1)F}},$$

where $n$ is the number of equal bins into which the $\cos\theta_{\nu e}$ distribution has been divided. As $n$ is made larger than 2, this expression quickly approaches $\sqrt{\frac{3}{F}}$. In Table II $B'$ and $\sigma(B')$ are given for $n \to \infty$ and one year's data. The anisotropy at a single energy is far smaller than the statistical uncertainty at current background levels. In a linear least-squares fit for the anisotropy of the expected size at all energies, the cumulative effect reaches one standard deviation (i.e., $\Sigma \left[B'/\sigma(B')\right]^2 = 1$) in approximately 30 years. Even with substantial reductions in the background, a decisive signal appears out of reach. The annual variation is also unobservably small.

The minor oxygen isotopes are also of interest in the heavy-water detector SNO [14]. In that case, where the backgrounds above 6.5 MeV are expected to be negligible, the principal effect is to augment the dominant event rate from deuterium. Some isotopic enrichment of $^{18}$O and $^{17}$O occurs in the production of heavy water; a sample has been assayed [15] at 0.334% $^{18}$O, but this may not be representative of the full 1000-tonne inventory. At that level, $^{18}$O would contribute 0.78% to the charged-current rate, an effect a factor 7 below the estimated theoretical uncertainty in the cross section on deuterium. The electron spectrum from neutrino interactions on deuterium is somewhat softer than in the $^{18}$O case as energy is given to the two protons in the continuum, but the modification to the spectral shape caused by $^{18}$O remains negligible compared to statistical uncertainties for any foreseeable data-taking period.

Although the 6.7% contribution of $^{18}$O and $^{17}$O to the solar neutrino signal appears appreciable in SuperKamiokande, we find that its effects remain hidden under current - and most likely future - background conditions. In fact, the isotropy of the cross section is so precise in the lower-energy bins that it simplifies certain proposed searches [16] for new physics. For example, some oscillation scenarios can produce active electron antineutrinos which, on interacting with protons in the water, generate a gentle anisotropy in the produced positrons. This anisotropy, which would be almost independent of $E_e$, clearly could not be attributed to the nuclear effects we have discussed.

This work was supported in part by the US Department of Energy.

Email addresses: haxton@phys.washington.edu, rghr@u.washington.edu

TABLES

TABLE I. Allowed nuclear matrix elements $|M|^2 = $ BGT + BF and the resulting cross sections, averaged over an undistorted $^8$B neutrino flux.

| Target | $E_f$(MeV) | BGT | BF | $\sigma(^8B)$ ($10^{-42}$ cm$^2$) |
|---|---|---|---|---|
| $^{18}$O | 0.0 | 5.12 | | 4.14 |
| | 1.04 | | 2.0 | 1.11 |
| | 1.70 | 0.21 | | 0.103 |
| total | | | | 5.35 |
| $^{17}$O | 0.0 | 1.69 | 1.0 | 1.53 |

TABLE II. Differential rates in units of counts/day/22.5 kilotons/MeV. All rates are for 1 MeV bins centered on $E_e$. Background rates are taken from Ref. [5].

| $E_e$(MeV) | R($^{18}$O+$^{17}$O+d) | $R(\nu-e)$ | $B(E_e)$ | Background | $B'(E_e)$ | $\sqrt{\frac{3}{F(E_e)}}$ | $R^{b-f}/R(\nu-e)$ |
|---|---|---|---|---|---|---|---|
| 5  | 0.176 | 10.86 | 0.001 |       |        |       | 0.1%  |
| 6  | 0.243 | 8.99  | 0.02  |       |        |       | 0.03% |
| 7  | 0.292 | 7.05  | 0.047 | ∼78   | 0.0002 | 0.004 | 0.10% |
| 8  | 0.307 | 5.17  | 0.073 | ∼47   | 0.0004 | 0.005 | 0.22% |
| 9  | 0.281 | 3.49  | 0.105 | ∼28   | 0.0009 | 0.006 | 0.42% |
| 10 | 0.213 | 2.12  | 0.150 | ∼14   | 0.0019 | 0.008 | 0.75% |
| 11 | 0.124 | 1.11  | 0.217 | ∼6.2  | 0.0035 | 0.012 | 1.22% |
| 12 | 0.047 | 0.46  | 0.305 | ∼2.8  | 0.0043 | 0.018 | 1.54% |
| 13 | 0.007 | 0.13  | 0.330 | ∼1.2  | 0.0017 | 0.028 | 0.95% |

FIGURES

FIG. 1. The differential capture rate dR/dcos$\theta_{\nu e}dE_e$ in counts/22.5 kilotons/day/Mev for $(\nu, e)$ reactions on $^{18}$O (dashed line) and the total (solid line) from $^{18}$O, $^{17}$O, and deuterium. A $^8$B neutrino flux reduced to 0.368 of the SSM value has been used.

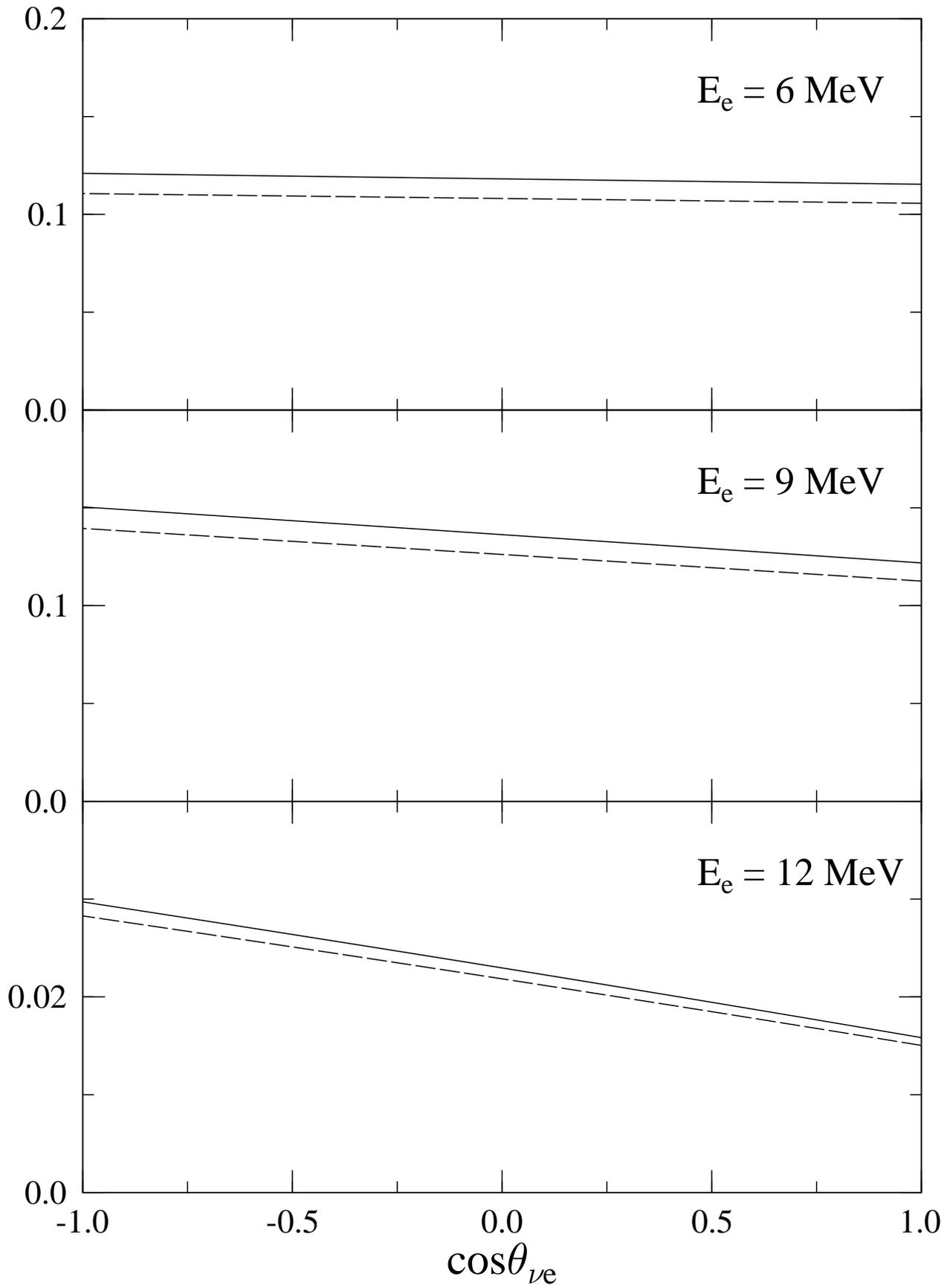